\documentstyle[prb,twocolumn,psfig,aps,epsf,graphics]{revtex}
\begin{document}
\draft
\twocolumn[\hsize\textwidth\columnwidth\hsize\csname
@twocolumnfalse\endcsname
%
%
%

\title{Relation between flux formation and pairing
in doped antiferromagnets}

\author{P. Wr\'obel$^{1}$ and R. Eder$^2$}

\address{$^1$ Institute for Low Temperature and Structure
Research,
P. 0. Box 1410, 50-950 Wroc{\l}aw 2, Poland}
\address{$^2$ Institut f\"ur Theoretische Physik, Universit\"at W\"urzburg,
Am Hubland, 97074 W\"urzburg, Germany}

\date{\today}
\maketitle

\begin{abstract}
We demonstrate that patterns formed by the current-current
correlation function are landmarks which indicate that spin
bipolarons form in doped antiferromagnets. Holes which constitute
a spin bipolaron reside at opposite ends of a line (string) formed
by the defects in the antiferromagnetic spin background. The
string is relatively highly mobile, because  the motion of a hole
at its end does not raise extensively the number of defects,
provided that the hole at the other end of the line follows along
the same track. Appropriate coherent combinations of string states
realize some irreducible representations of the point group
$C_{4v}$. Creep of strings favors $d$- and $p$-wave states. Some
more subtle processes decide the symmetry of pairing. The pattern
of the current correlation function, that defines the structure of
flux, emerges from motion of holes at string ends and coherence
factors with which string states appear in the wave function of
the bound state. Condensation of bipolarons and phase coherence
between them puts to infinity the correlation length of the
current correlation function and establishes the flux in the
system.
\end{abstract}


\vskip2pc] \narrowtext

\section{Introduction}
Flux phase  appeared in the mean-field approaches to undoped and
doped antiferromagnets described by the $t$-$J$ model\cite{Lee89}.
The physical meaning of that phase and its relation with pairing
was not clear. In some announcements
\cite{Fu,Dong,Ziqiang,Ubbens,Sandalov,Cappelluti} the competition
or coexistence between superconductivity and the flux phase have
been discussed in the framework of the mean filed approach and the
$1/N$ expansion. Some researchers tend to understand the spin gap
phase in doped antiferromagnets as the staggered flux phase
\cite{Wen,Lee}. Other believe that pairing grows out of the
$\pi$-flux phase \cite{Dong} or that pairs are formed by holes
circulating in opposite directions \cite{Ivanov00}. In this paper
we will present a different point of view that the pattern of
currents and pairing originate in formation of spin bipolarons of
a particular form. We will give some mathematical arguments, based
on understanding of pairing in weakly doped antiferromagnets, in
support of that idea.

 Recently, Ivanov, Lee and Wen \cite{Ivanov00} reported
staggered-vorticity correlations of the current-current
correlation function $\langle j_{ij} j_{kl}\rangle $ (CCCF) in the
$d$-wave variational functions for a weakly doped antiferromagnet,
where $j_{ij}$ denotes the current flowing on the bond $\langle
i,j \rangle$. Similar pattern of the CCCF was found for a bound
state of two holes in an exact diagonalization performed by Leung
\cite{Leung}. Since the mechanism of hole binding in weakly doped
antiferromagnets is relatively well understood
\cite{WrobelEder98}, it is a tempting task to resolve whether
there exist a deep relation between binding and the formation of
flux patterns.

Fast motion, with the rate $\sim t$, of a hole created at some
site in the N\`eel state forms defects in the original arrangement
of spins \cite{BulaevskiiNagaevKhomskii,ShraimanSiggia}. These
defects, that are spins which have been turned upside down and may
also be called magnons, lie on the track of the hole. Strings
pinned to a central site form a potential well for a hole which
confines its motion in the vicinity of this site. Much slower
processes related to the inversion of anti-parallel spins at
nearest neighbor sites which occur at the rate $\sim J$ shorten
strings and give rise to coherent propagation of the hole. Since
the separation  of energy scales for hopping of a hole and
annihilation of defects (magnons) is pronounced it is plausible to
introduce the notion of a spin polaron. We may define in this
context a spin polaron as a solution of the Schr\"odinger equation
for a particle in the potential well \cite{EderBecker}. Slower
processes are neglected at this stage of analysis. ``Orbital''
states of polarons created at all possible sites exhaust, in
principle, the relevant portion of the Hilbert space. In practice,
the calculation may be confined to some low excited states or even
to the polaron groundstate. By calculating the matrix elements of
the full Hamiltonian in the polaron basis we derive an effective
Hamiltonian expressed in the polaron language. Some processes,
like oscillations of the hole in the vicinity of a polaron center,
are already incorporated into the eigenenergy of polaronic states.
The rest either renormalizes the eigenenergies or gives rise to
off-diagonal matrix elements in the polaronic Hamiltonian.  The
latter circumstance occurs  in the case of shortening of strings
by an appropriate term in the original Hamiltonian. The approach
outlined above has some obvious limitations. It is applicable
provided that the correlation length of background
antiferromagnetic spin correlations is larger than the polaron
radius.

We extend now the concept of spin polarons to the description of
interaction between holes \cite{Eder}.  Since we are going to
apply this construction to the calculation of the current
correlation function, we will  provide the reader with more
details. We define a spin bipolaron $|\Psi_{\langle i,j \rangle
}\rangle$ as a combination of states which may be reached by
independent hopping of holes created at a pair of nearest neighbor
sites $\langle i,j \rangle$,
\begin{equation}
|\Psi_{\langle i,j \rangle }\rangle = \sum_{{\cal P}_i,{\cal P}_j}
\alpha_{{\cal P}_i,{\cal P}_j} |{\cal P}_i,{\cal P}_j\rangle.
\label{wvfn}
\end{equation}
${\cal P}_i$ parameterizes the geometry of a path along which the
hole has been moving and $|{\cal P}_i,{\cal P}_j\rangle$ is a
state which has been created in this way.  Due to Fermi
statistics, order in which holes have been originally created in
$|\Psi_{\langle i,j \rangle
  }\rangle$ is relevant. We adopt a convention, according to which
a hole is created first at the site $i$ and  $i$
belongs to the even sublattice. At this stage of
considerations we prohibit, by definition, each
hole to follow along the trace left by the accompanying hole. By means
of that somewhat artificial restriction we achieve that spin
bipolarons are localized and we may proceed as in the case of a single
hole. We make a further approximation and neglect all path details
including the possibility of path crossing. It is self-evident
that the coefficients $\alpha_{{\cal P}_i,{\cal P}_j}$ for a
bipolaron in the groundstate will  depend only on the lengths $\mu$,
$\nu$ of paths ${\cal P}_i$ and ${\cal P}_j$,  $\alpha_{{\cal
  P}_i,{\cal P}_j}=\alpha_{\mu,\nu}$. The Schr\"odinger equation that
describes  a hole pair in a potential well and defines the spin bipolaron,
may be written as
\begin{eqnarray}
t \left[\alpha_{\mu-1,\nu} + (z-1) t \alpha_{\mu+1,\nu} +
\alpha_{\mu,\nu-1} + (z-1) t \alpha_{\mu,\nu+1}\right] \nonumber \\
+
J\left(4+\mu+\nu-\frac{1}{2}\delta_{\mu,\nu}\right)
\alpha_{\mu,\nu}=E_2
\alpha_{\mu,\nu},
\end{eqnarray}
where $\alpha_{\mu,\nu}=0$ for $\mu<0$ or $\nu<0$ and z=4. The
form of this equation is easy to understand. The first expression
in it is related to the fact that each path may be reached from
one shorter path and (z-1) different longer paths by a hop of the
hole. The second term counts the number of pairs of nearest
neighbor sites, that are not occupied by anti-parallel spins.
Every such a ``broken bond'' raises the energy by $J/2$ in
comparison to the energy of the N\'eel state.

Formation of bipolarons, and, in particular, strings that connect
holes, is an effective way of lowering energy. A compromise
between two opposite tendencies to minimize the kinetic energy of
holes and to reduce the disturbance of the antiferromagnetic
background is reached by means of that process. By shrinking at
one end and expanding at the opposite end, a string may move,
while keeping a moderate length \cite{Trugman}. The application of
the spin-polaron scenario to binding of holes in doped
antiferromagnets gave rise to better than qualitative agreement
with results of numerical diagonalizations \cite{WrobelEder}. In
particular the hierarchy of states that realize irreducible
representations for all momenta and symmetries allowed by the
geometry of the $4 \times 4 $ cluster have been reproduced
\cite{WrobelEder98}. It turns out that creep of strings favors
$d$-wave and $p$-wave states. More subtle processes decide the
symmetry of the bound state in favor of the $d$-wave. Some
modifications of the $t$-$J$ model, like asymmetry of the magnetic
interaction may lead to stability of the $p$-wave state
\cite{ChernyshevLeung}. The weight of bipolaronic states in the
bound state amounts to about $80 \%$ in the case of the $d$-wave
and $50 \%$ in the case of the $p$-wave \cite{WrobelEder98}.  Thus
it is apparent that in a qualitative description of current
correlation function in the presence of pairing we may neglect
monopolaronic states.

$s$-, $d$- and $p$-wave symmetries may
be realized as coherent sums of bipolaronic states
\cite{WrobelEder98}: $\sum_{\langle i,j \rangle } S_{\langle i,j
  \rangle} | \Psi_{\langle i,j \rangle}\rangle$, where  $S_{\langle i,j
  \rangle}=1$ if  $\langle i,j \rangle$ is horizontal
and $S_{\langle i,j \rangle}=-1$ provided that  $\langle i,j \rangle$ is
vertical for the $d$-wave, while for the $p$-wave $S_{\langle i,j
  \rangle}$ vanishes if  $\langle i,j \rangle$ is vertical,
$S_{\langle i,j \rangle}=1$ provided that  $j$ is on the right
side of $i$ and $S_{\langle i,j \rangle}=-1$ if on the left. In
order to make the story short we have skipped many details. The
full construction of bipolarons that conform with  irreducible
representations of the point group $C_{4v}$ may be found in an
earlier article \cite{WrobelEder98}.

\section{Flux pattern of the $d$-wave paired state}
As a first example of a contribution to the CCCF we consider a
process shown in Fig.~\ref{str1}(a). The middle row and the lowest
row depict string states of the type $|{\cal P}_i,{\cal
P}_j\rangle$ which are coupled by action of a product of current
operators on outer bonds. Circles represents positions of
bipolarons, or in other words sites at which holes have been
created, while thin arrows symbolize paths. In the case of the
state in the second row, a single defect in the N\`eel structure
occupies the second site from the left, while in the case of the
final state in the third row a single defect (magnon) occupies the
third site from the left. Current operators which are given by the
formula $j_{ij} \propto i (c^{\dag}_{i\sigma} c_{j\sigma}-
c^{\dag}_{j\sigma} c_{i\sigma})$ shift holes in the middle row.
 Holes are moved in the same direction,
which is represented by the arrows in the uppermost row.
\begin{figure}[htbp]
\unitlength1cm
\begin{picture}(8.0,1.9)
\epsfxsize=8.2cm \put(0,0.3){\epsfbox{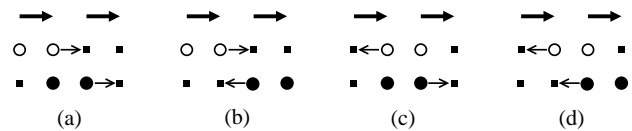}}
\end{picture}
\caption{Some processes which contribute to the CCCF
  function when holes are shifted by a product of current operators in the same direction on outer bonds.}
\label{str1}
\end{figure}
 If we adopt
a convention that the direction of current on both outer bonds is
from left to right the contribution to the CCCF is $t^2
\alpha_{0,1}^2$. The factor $i^2$ which stems from product of two
current operators has been absorbed by a change of sign related to
the fact that the hole created at the even site in the middle row
landed at a site which is occupied in the lowest row by a hole
created at the odd site. Fig.~\ref{str1}(a) is an example of a
contribution to a the CCCF for two bonds which may be connected by
a line consisting of a single bond. If we incorporate the outer
bonds to that line the length of it amounts to three. Since we
expect that the absolute value of amplitudes $\alpha_{\mu,\nu}$
decreases when the total length of two involved paths $\mu+\nu$
increases, only the shortest possible string states will decide
the sign of the CCCF for two chosen bonds.
Figures \ref{str1} (b), (c), (d) depict the rest of processes
which involve shortest paths for holes which are moved from left
to right at outer bonds of a line which length is $3$.  The
contributions from these processes to the CCCF are $-t^2
\alpha_{0,1}^2$, $-t^2 \alpha_{0,1}^2$, and $t^2 \alpha_{0,1}^2$
respectively. In the final formula these values are multiplied by
factors $S_{\langle i,j \rangle}$ with which bipolaronic states at
points represented by open and solid circles appear in the
coherent sum that defines the bound state. Analogous processes
contribute to the CCCF if holes are shifted from right to left on
both outer bonds.

\begin{figure}[htbp]
\unitlength1cm
\begin{picture}(8.0,2.5)
\epsfxsize=8.2cm \put(0,0.3){\epsfbox{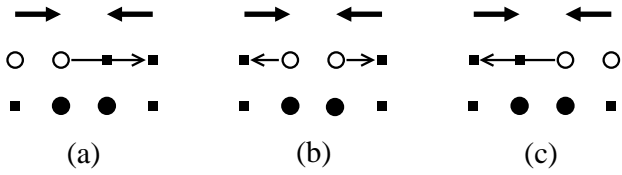}}
\end{picture}
\caption{Some processes which contribute to the CCCF
  when holes are moved by a product of current operators  in  opposite directions on outer bonds.}
\label{str2}
\end{figure}

\begin{figure}[htbp]
\unitlength1cm
\begin{picture}(8.0,5.7)
\epsfxsize=6.0cm \put(1.0,0.3){\epsfbox{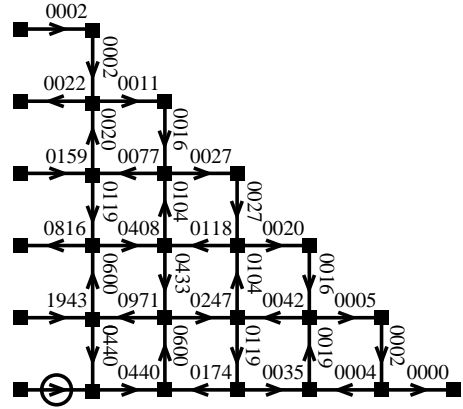}}
\end{picture}
\caption{ $\langle j_{kl} j_{mn} \rangle /x$ in units $10^{-5}$,
 where x is the hole
  concentration, at $J= 0.3 t$. The reference bond has been marked by a
  circle.}
\label{currcorr}
\end{figure}

\begin{figure}[htbp]
\unitlength1cm
\begin{picture}(8.0,3.2)
\epsfxsize=8.5cm \put(0.5,-3.3){\epsfbox{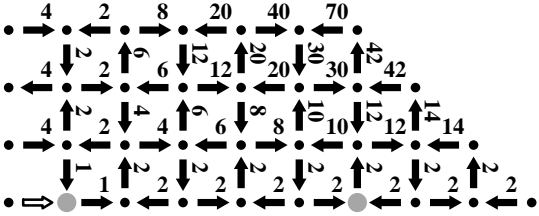}}
\end{picture}
\caption{The sum of expressions (\ref{currcorr}) for all shortest paths
  that connect bonds with a bond marked by an open arrow for the $d$-wave
  symmetry. The weights
  $\alpha$ have been neglected in the
  evaluation.}
\label{dxypatt}
\end{figure}

Fig.~\ref{str2} depicts contributions to the CCCF for holes which
move inward on outer bonds of a line of length $3$.
In the original string state, in the second row of Fig.~\ref{str2}
(a), to which the product of current operators is applied, holes
occupy outer sites, while two magnons reside on a pair of sites in
the middle. These magnons (defects) disappear when holes are
shifted to the central position in the lowest row. Since holes
that have been created in even and odd sublattices, exchange
positions, the Fermi statistics again contributes the factor
"$-1$" but holes are moved in opposite directions, and the
contribution to the CCCF becomes $-t^2 \alpha_{0,0} \alpha_{0,2}$.
Amplitudes of processes in Figures \ref{str2} (b) and (c) are $
t^2 \alpha_{0,0} \alpha_{1,1}$ and $-t^2 \alpha_{0,0}
\alpha_{0,2}$ respectively. The directions of hole motion  may be
inverted if the product of current operators is applied to the
string state in the lowest row which gives rise to next three
contributions to the CCCF on outer bonds. We can use the insights
gained from the analysis of a line which consists of $3$ links in
total, to obtain a formula for the contribution from bipolaronic
states which lie on a line of arbitrary length $l$, to the CCCF on
outer bonds of that line,
\begin{eqnarray}
2 \sum_{m=1}^{l-1}  \sum_{n=2}^{l} (-1)^{m+n+1} S_m S_n
t^2 \alpha_{m-1,l-m-1} \alpha_{n-2,l-n} \nonumber \\
+  2 \sum_{m=1}^{l}  \sum_{n=2}^{l-1} (-1)^{m+n} S_m S_n
t^2 \alpha_{m-1,l-m} \alpha_{n-2,l-n-1},
\label{ccf}
\end{eqnarray}
where $S_m$ is a coherence factor with which the bipolaron at
$m$-th bond in the line appears in a coherent sum that represents a
bound state.
 An additional lesson of the
above analysis is that the shortest string states which contribute
with the highest weight to the CCCF for a pair of bonds, lie on
shortest paths that lead through lattice points and incorporate
these bonds. If the length of such paths amounts to $l$, the
length of involved strings is $l-1$, when holes are moved by a
product of current operators in the same direction or $l$ and
$l-2$, when the holes are shifted in opposite directions. As we
have already mentioned, the leading contribution to the CCCF for a
pair of chosen bonds is the sum of the expression (\ref{ccf}) over
all shortest paths that incorporate those bonds.
Fig.~\ref{currcorr} represents in units $10^{-5}$ the CCCF divided
by the hole concentration, that has been calculated in this way
for the d-wave bound state of two holes. Arrows represent the
direction of current on respective bonds. We notice that despite
the crudeness of the applied approximation, agreement with results
of exact diagonalization performed by Leung \cite{Leung} is much
better than qualitative. It is an intriguing question, if that
coincidence is a manifestation of a deeper relation between
pairing and patterns of the CCCF. In order to see that this is the
case we assume that the logarithm of the weight of a string
$\alpha$ is proportional to its length $l$. That assumption is
very natural because  by expanding the string by one lattice
spacing we raise the number of broken bonds by 2 and reach 6
longer string states, independently of the value of $l$. In our
description of spin polarons parameters $\alpha_{\mu,\nu}$, where
$\mu+\nu=l-1$, are related with strings of length $l$, but in
general $\alpha_{\mu,\nu} \neq \alpha_{\mu^\prime,\nu^\prime}$ for
$\mu+\nu=\mu^\prime+\nu^\prime$, which is the prize which we pay
for pinning strings and making bipolarons localized.
\noindent We notice that the total length of strings contributing
to elements of sums in (\ref{ccf}) is $2l-2$. Under the previously
made assumption, that  $\alpha_{\mu,\nu} \propto \exp(C
(\mu+\nu))$,  all products of two weights $\alpha$ are the same in
both sums. If we neglect them and evaluate the sum of expressions
for all shortest lines that link bonds we again get the staggered
flux pattern presented in Fig.~\ref{dxypatt}. We notice that the
CCCF defined in this way has the expected structure and apart from
bonds attached to the reference bond the CCCF is given by twice
the number of shortest paths that connect bonds with the reference
bond. This is not by coincidence, because the sum (\ref{ccf}) with
factors $\alpha$ neglected, may be evaluated explicitly,
\begin{eqnarray}
2\sum_{m=1}^{l-1}  \sum_{n=2}^{l} (-1)^{m+n+1} S_m S_n \nonumber \\
+  2 \sum_{m=1}^{l}  \sum_{n=2}^{l-1} (-1)^{m+n} S_m S_n
=(-1)^l 2 S_1 S_l.
\label{splccf}
\end{eqnarray}
This formula implies that the pattern of the CCCF is determined in a
simple way by the symmetry of the bound state or to be more specific
by the coherence factors with which bipolarons on involved bonds
appear in the wave function.

\section{Density correlation function}
\begin{figure}[htbp]
\unitlength1cm
\begin{picture}(8.0,4.2)
\epsfxsize=8.5cm \put(-1.0,-7.1){\epsfbox{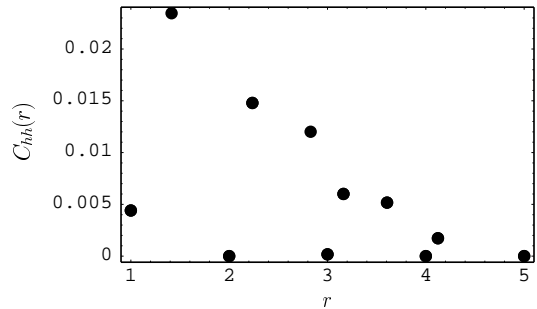}}
\end{picture}
\caption{Contributions to the DDCF, $C_{hh}(r)$ at $J=0.3t$ from
  string states that lie on
  shortest paths which connect relevant sites.}
\label{dencorr}
\end{figure}

There exists another quantity which behavior is determined by
existence of spin bipolarons. Density matrix renormalization group
calculations \cite{WhiteScalapino} and exact diagonalizations
\cite{Leung} point out that the density-density correlation
function (DDCF) $C_{hh}(r)$ for holes as a function of distance
$r$  between them reveals a characteristic structure in the d-wave
bound state. The most pronounced feature of it is that the DDCF
decays much faster along lines in the $x$ ($y$) direction,
especially if the distance $r$ is an even number. We assume again
that the highest contributions to the DDCF stem from shortest
possible string states. These strings should stretch between
relevant sites in order to contribute to the DDCF. Thus, the decay
of the DDCF with distance is related to the decay of CCCF by the
dependence of weights parameters $\alpha$ on string length. Gray
disks in Fig.~\ref{dxypatt} point out to the relation between the
DDCF at a pair of sites and CCCF on bonds attached to them. After
a short evaluation, the contribution to the DDCF from strings that
lie on a line takes the form,
\begin{equation}
2 \sum_{m,n=1}^{l}(-1)^{m+n} S_m S_n
\alpha_{m-1,l-m} \alpha_{n-1,l-n}.
\label{ddcf}
\end{equation}
For the $d$-wave state that sum exactly vanishes for points at one
of the axis ($x$ or $y$) if $l$ is even. Fig.~\ref{dencorr}
depicts the DDCF obtained for the $d$-wave state by adding
contributions (\ref{ddcf}) from all shortest lines that connect
relevant sites.
Notwithstanding the simplicity of our approach the similarity with
results of numerical diagonalization by Leung \cite{Leung} is
evident. Oscillations of $C_{hh}(r)$ with distance and minima at
points which lie on an axis are visible in
both calculations. Thus, we  draw a conclusion that the pattern of the
CCCF, the structure of the DDCF and the relation between rates of
their decay with distance in the $d$-wave state is related with
formation of bipolarons and strings.

\section{Flux pattern of the $p$-wave paired state}
\begin{figure}[htbp]
\unitlength1cm
\begin{picture}(8.0,3.2)
\epsfxsize=7.5cm \put(0.5,-3.0){\epsfbox{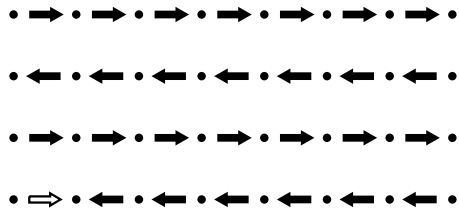}}
\end{picture}
\caption{The pattern of currents in the CCCF for the $p_x$-wave
symmetry.} \label{pxpatt}
\end{figure}

 Motivated by this observation we may try to guess the
pattern of CCCF in the $p$-wave bound state. A state with that
symmetry may became a groundstate if the exchange interaction
between spins in the $t$-$J$ model becomes anisotropic. By
applying the formula (\ref{splccf}) we get the structure presented
in Fig.~\ref{pxpatt} for the case of the symmetry $p_x$.
 \noindent No
current flows in the vertical direction because there are no
bipolarons on vertical bonds. The cancellation which lead to the
formula (\ref{splccf}) is not exact in the case of the expression
(\ref{ccf}), and in addition some higher order contribution to the
CCCF from spin bipolarons will appear for longer paths. Evaluation
of these contributions to the CCCF is beyond the scope of this
paper. Monopolarons may also obscure the picture presented in
Fig.~\ref{pxpatt} because the weight of bipolarons in the $p$-wave
state is smaller than for the $d$-wave state.

\section{Conclusions}
In summary, we have demonstrated that the vorticity correlations
in weakly doped antiferromagnets have their origin in formation of
spin bipolarons and strings. The flux pattern is determined by the
symmetry of the bound state. The rate of decay with distance of
different correlation functions  related to current or density is
determined by  the rate with which the weight of strings
diminishes with their length. That observation indicates that
pairing and vorticity correlations in doped antiferromagnets have
common origin in formation of bipolarons or in other words of
strings built by magnons that connect holes at their endpoints.

The proximity of both states war already observed in mean field
calculations. Conclusions drawn from that observation that pairing
grows out of flux or circulating currents are questionable in the
light casted by calculations presented in previous sections. Also
the statement that pairs are formed by holes that circulate in
opposite directions does not agree with our scenario. It seems
that agreement between analytical calculations and exact
diagonalizations suggest that holes that form a bound state
oscillate in all possible directions at opposite endpoints of a
string.

Depletion of the single particle spectral weight around the Fermi
level \cite{Timusk} known as pseudogap and exceptional
non-Fermi-liquid (NFL)properties of copper-oxide based metals
\cite{Alloul,Bucher} are characteristic features of those systems
that decide about the complexity of their physics. Some
suggestions have been expressed that phase fluctuations are
responsible for the pseudogap phenomenon \cite{Emery}. These ideas
are tightly related with the Bose-Einstein condensation of local
pairs \cite{Uemura}. The microscopic derivation in the framework
of a two band model of the marginal Fermi liquid (MFL) concept
\cite{Varma} which seems to describe quite well the
 NFL behavior observed in  copper oxides is based
on existence of circulating currents. The calculation presented in
this paper suggests that both local pairs responsible for
pseudogap phenomena and circulating currents which give rise to
the MFL behavior may be attributed to formation of spin
bipolarons. Finally, it is important to emphasize that both
phenomena are tightly related and can not be discussed separately
in the case of the copper oxides, which is also in agreement with
the suggestion their relation with spin bipolarons and string
states. \vspace{0.3cm}

 One of the authors (P.W.) acknowledges support by the
Polish Science Committee (KBN) under contract No. 5 P03B 058 20.

\end{document}